\def\year{2022}\relax
\documentclass[letterpaper]{article} 
\usepackage{aaai22}  
\usepackage{times}  
\usepackage{helvet}  
\usepackage{courier}  
\usepackage[hyphens]{url}  
\usepackage{graphicx} 
\usepackage{natbib}  
\usepackage{caption} 
\DeclareCaptionStyle{ruled}{labelfont=normalfont,labelsep=colon,strut=off} 
\usepackage{algorithm}
\usepackage{algorithmic}
\usepackage{tabularx}
\newcolumntype{s}{>{\hsize=.45\hsize}X}
\usepackage{newfloat}
\usepackage{listings}
\floatstyle{ruled}
\newfloat{listing}{tb}{lst}{}
\floatname{listing}{Listing}

\usepackage{amsmath, amssymb, amsfonts}
\usepackage{multirow}
\usepackage{cleveref}
\usepackage{enumitem}

\title{Susceptibility of Communities against Low-Credibility Content in Social News Websites}
\author {
    Yigit Ege Bayiz,\textsuperscript{\rm 1}
    Arash Amini,\textsuperscript{\rm 2}
    Radu Marculescu,\textsuperscript{\rm 1}
    Ufuk Topcu\textsuperscript{\rm 2}
}
\affiliations {
    \textsuperscript{\rm 1} Department of Electrical and Computer Engineering, The University of Texas at Austin\\
    egebayiz@utexas.edu, radum@utexas.edu
    
    \textsuperscript{\rm 2} Oden Institute for Computational Engineering and Sciences, The University of Texas at Austin\\
    a.amini@utexas.edu, utopcu@utexas.edu

}

\newcommand{\DRed}{\textbf{Reddit Dataset}}
\newcommand{\DAgg}{\textbf{(Dis)agreement Dataset}}
\newcommand{\DNeg}{\textbf{Negation Dataset}}

\newcommand{\dred}{\textbf{Reddit dataset}}
\newcommand{\dagg}{\textbf{(dis)agreement dataset}}
\newcommand{\dneg}{\textbf{negation dataset}}

\begin{document}
\maketitle

\begin{abstract}
    Social news websites, such as Reddit, have evolved into prominent platforms for sharing and discussing news. A key issue on social news websites sites is the formation of echo chambers, which often lead to the spread of highly biased or uncredible news. We develop a method to identify communities within a social news website that are prone to uncredible or highly biased news. We employ a user embedding pipeline that detects user communities based on their stances towards posts and news sources. We then project each community onto a credibility-bias space and analyze the distributional characteristics of each projected community to identify those that have a high risk of adopting beliefs with low credibility or high bias. This approach also enables the prediction of individual users' susceptibility to low credibility content, based on their community affiliation. Our experiments show that latent space clusters effectively indicate the credibility and bias levels of their users, with significant differences observed across clusters---a $34\%$ difference in the users' susceptibility to low-credibility content and a  $8.3\%$ difference in the users' susceptibility to high political bias.
\end{abstract}

\section{Introduction}
\textit{Social news websites}, such as Reddit and Digg, have emerged as primary platforms for exchanging, archiving, and accessing information. These platforms enable users to share opinions and news articles, and provide an open forum where their users can comment on, discuss, or criticize the news. Their ability to allow news sharing and discussion with minimal censorship allowed social news websites to flourish as open repositories of news from diverse sources and opinions. Social news websites have become a common way for people to access news content. A $2023$ report by the Pew Research Center indicates that $8\%$ of U.S. adults \textit{regularly} rely on Reddit for news \cite{PewResearch-2023-FactSheet}.


The openness of social news websites also serves as a fertile ground for the spread of uncredible or highly biased information. A notable example is r/politics on Reddit, the largest political news discussion community, where more than half of the shared sources contain unverifiable content. The prevalence of unverified news, amplified by the content recommendation algorithms of these sites, tends to reinforce and strengthen users' pre-existing beliefs. This phenomenon leads to significant exposure to news with uncredible or highly biased origins among some user communities. Such communities play a substantial role in propagating uncredible or biased narratives, potentially causing a spectrum of social issues ranging from creating confusion and distracting users from correct news, to leading people to support extremist or hyper-partisan beliefs. 


{\renewcommand{\arraystretch}{1.15}
\begin{table}[t]
    \centering
    \begin{tabularx}{230pt}{X|ccc}
        \hline
        
        \textbf{Subreddit}& \textbf{\# Ver.} & \textbf{\# Unver.} & \textbf{\% Unver.} \\
        
        \hline
        r/Conservative & $37,593$ &  $64,195$& 72\%  \\
        r/Libertarian & $15,366$ &  $83,618$& 16\% \\
        r/democrats & $5,875$ &  $12,076$& 77\% \\
        r/Republican & $12,943$ & $19,129$& 72\% \\
        r/politics & $598,844$ & $642,634$& 52\% \\
        total &$670,621$&$821,652$& 55\%\\
        \hline
    \end{tabularx}
    \caption{Comparison of numbers of verifiable (Ver.) and unverifiable (Unver.) submissions over the five largest political subreddits in Reddit.}\label{Table:subreddits_ver_ratio}
\end{table}}

Detecting and countering uncredible or highly biased news content is a well-researched problem. Numerous deep learning methods have been developed to identify such news sources, as highlighted in studies  \cite{Zhou2020FakeNews, monti2019fake}. Additionally, there's a growing trend to employ so-called large language models for this purpose \cite{hu2023llmmisinfo}. Efforts also extend to identifying major spreaders of uncredible content among users \cite{sakketou2022factoid}. Such efforts aim to detect the optimal targets for preventative methods such as moderation and banning. 


This paper adopts a novel perspective by focusing on the detection of \textit{communities} with a high susceptibility to uncredible or highly biased news. In this context, we define a community as a substantial group of users sharing similar opinions, interests, or beliefs, and exhibiting similar reactions to news articles. We propose a novel comment-based user embedding methodology to create latent space embeddings for individual users and we investigate the relation of these embeddings with users' susceptibility to interact positively with uncredibile or highly biased news content. Specifically, we utilize our embedding method to cluster users into communities. We then analyze the distribution of credibility and political biases of these communities.


Pretrained sentence embedding models like sentence-BERT (SBERT) \cite{reimers-2019-sentence-bert} have significantly advanced the embedding of social media content, enabling research on content clustering and analysis to discern user opinions and biases. However, there is no consensus on inferring user opinion embeddings from the content they engage with. One method involves pooling user-posted content to average the embeddings of each post. This method, though straightforward, is impractical due to the insufficient volume of posts per user for reliable embedding estimation. 


We address these challenges by deriving user embeddings from user comments rather than the shared news sources directly. This approach yields a larger data set from users, reducing statistical variance in latent space representations. We provide context to the user comments based on their stances towards the original news post, and use this contextual information to assign embeddings to the comments. Then, we use averaged pooling on the comment embeddings to gather user embeddings. This method ensures that the user embeddings reflect user opinions and interests on a similar latent space to the post embeddings.

We apply our embedding method to real-world data from Reddit, a social news platform with user-generated interest groups called subreddits. On Reddit, users can post opinions or news, and engage with others through comments and replies. After embedding users, we identify user communities and examine their credibility and bias distributions. Our goal is to answer the following research questions.

\begin{itemize}[font=\bfseries,align=left]
    \item[\textbf{RQ1:}]  How do user communities in Reddit differ in their susceptibility to credibility and bias?
    \item[\textbf{RQ2:}] Is it possible to predict the likelihood of a user responding positively to low-credibility news, based on their cluster assignment?
\end{itemize}

Determining the credibility and biases of news sources is often subject to the individual biases of those who rate them. In this paper, we rely on the data released by Ad Fontes Media\footnote{This dataset is detailed at \url{https://adfontesmedia.com}}, a public benefit corporation that aims to counter misinformation and highly biased media. This dataset includes credibility and bias scores of $223$ news sources. We use this data to assign credibility and bias scores to Reddit posts that reference one of these news sources. We call such posts \textit{verifiable} and likewise, if a post does not contain a news link contained in this dataset, we call that post \textit{unverifiable}. \Cref{Table:subreddits_ver_ratio} provides a breakdown of the number of verifiable posts in five major political subreddits in Reddit.

\section{Related Work}
\subsection{Sentence Embedding}
Sentence embedding is a critical invention that enables automated analysis of social news content. These models work by assigning a numerical representation of each sentence that preserves the syntactic and semantic relation between sentences. Early approaches to sentence embedding models involve encoder-decoder architectures such as Skip-thought \cite{Kiros2015sembed}, and LSTM-based structures such as siamese BiLSTM \cite{conneau-etal-2017-supervised}. 

Modern sentence embedding relies on using pre-trained transformer-based architectures \cite{Vaswani2017Bert}. Chief among them is the Bidirectional Encoder Representations from Transformers (BERT) \cite{devlin2019bert}, which set state-of-the-art performance in semantic textual similarity benchmark. Later, RoBERTa \cite{liu2019roberta}, improved this benchmark performance by utilizing small optimizations in BERT pre-training. 

Intrinsically, both BERT and RoBERTa are incapable of achieving sentence embedding as they do not derive independent sentence embeddings. \citet{reimers-2019-sentence-bert} enabled drawing such sentence embeddings by introducing sentence-BERT (SBERT) which incorporates a pooling layer after the pre-trained BERT network, and train them using the siamese network architecture in which they train two copies of the same network simultaneously on a sentence similarity or classification objective. In this paper, we use this SBERT architecture for embedding Reddit posts. 


\subsection{Stance Detection}
In this paper we use \textit{stance detection} as a part of our user embedding pipeline. Stance detection entails classifying the sentiment of a text, such as a user comment, towards a given target, \cite{Kucuk2020std}. Stance detection was pioneered by \citet{qazvinian-etal-2011-rumor}. Later \citet{augenstein-etal-2016-stance} achieved state-of-the-art performance by utilizing bidirectional encoding architectures. Modern stance detection relies mostly on transformer models \cite{hardalov-etal-2022-survey}, with  \citet{arakelyan-etal-2023-topic} achieving state-of-the-art performance.

Recently, \citet{pougue-biyong2021debagreement} curated a comment-reply dataset with stance labels collected over Reddit. We use this dataset to train and validate our stance detection method, a variant of the method proposed by \citet{anon2023stanceDetection}. Recently \citet{luo2023daggresult} achieved state-of-the-art performance on this dataset, providing a baseline to compare our results.  

\subsection{User Profiling}
User profiling is the task of assigning a virtual representation to each user, such are keywords, personal information, or numerical latent space representations \cite{Eke2019UserProfiling}. Utilizing user profiling as a detection mechanism for fake news is not a new problem. \citet{Shu2018Profiling} study the statistical distributions on which Twitter\footnote{Accessible from \url{https://twitter.com/}} users are more likely to trust in false news based on gender, age, and personality traits. In a follow-up paper \cite{shu2020therole}, the authors extend upon these methods and introduce \textit{user profile features}, which is a high dimensional user representation that includes location, profile picture, and political bias information, and show that these features, in conjunction with text analysis methods such as linguistic inquiry word-count \cite{Pennebaker2015LIWC} and rhetorical structure theory \cite{ji-eisenstein-2014-representation} yield high classification accuracy and recall for false news detection. More recently, \citet{sakketou2022factoid} achieved state-of-the-art performance in detecting fake news sources by modeling the social interactions between Reddit users with a \textit{graph}, and using graph neural networks to classify nodes that are likely to spread misinformation \cite{Wu2021gnn}. Specifically, they construct a user-to-user graph by traversing the comments under Reddit posts and then use a graph attention network to classify fake news spreaders. 

This paper differs from the existing user profiling works in two regards. Firstly, while we develop embedding methods to gather high-dimensional representations of users based on their comments, we focus on extracting user communities from the high-dimensional user representations, rather than analyzing individual users. Secondly, contrary to existing studies we do not measure false news spreading probabilities. Rather, we characterize how user communities, characterized by their long-term commenting behavior, show differences in engaging with news from uncredible, or highly politically biased sources. 

\subsection{Contributions}
\begin{itemize}
    \item We introduce a user embedding pipeline that jointly uses stance detection, together with sentence encoders to obtain latent space representations of users
    \item We show that Reddit users create identifiable user communities based on their user embeddings.
    \item We show that the said communities indicate users' susceptibility to uncredible and highly biased news.
\end{itemize}

\section{Methods}

\subsection{User Embedding}
\begin{figure*}[t]
\centering
\includegraphics[width=\textwidth]{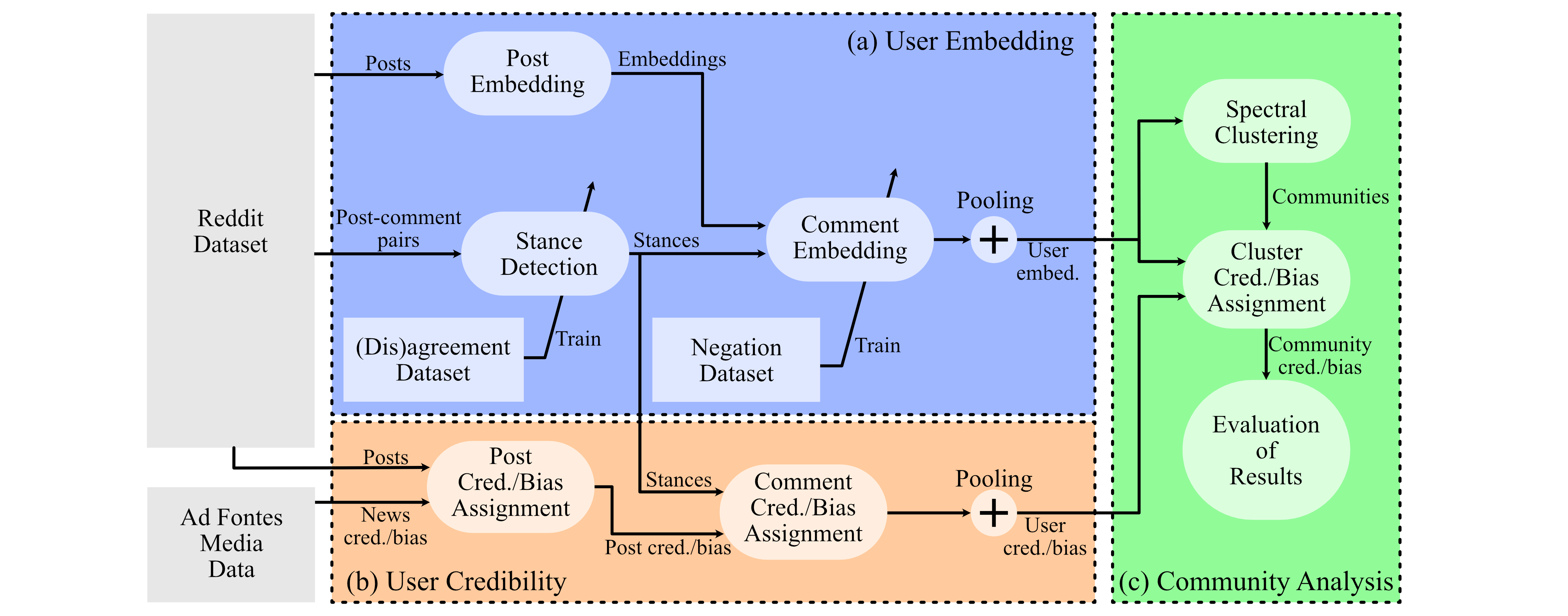} 
\caption{Overview of methodology: This diagram illustrates our analysis approach. Rounded boxes depict various processes, while sharp-edged boxes depict datasets. Diagonally upward arrows behind processes indicate the use of corresponding datasets for training these processes. The pooling blocks, indicated with a $\boldsymbol{+}$ sign denote \textit{averaging} the inputs for each user.}
\label{fig2}
\end{figure*}

In this section, we describe a method to embed the users in a high-dimensional latent space. This method works by first assigning an SBERT sentence embedding to posts, and then assign an embedding representation to comments by considering the stance of the comment towards the post. Then, we average the embeddings of all comments sent by each user to obtain a single latent space embedding of each user. This embedding representation captures the average interest and opinions of each user and their stances towards different viewpoints. \Cref{fig2}.a shows the overall structure of the user embedding process. In the following sections, we break down each element in this embedding process in more detail.


\subsubsection{Post Embedding} 
We embed the entire corpus of post titles using a pre-trained SBERT model, a sentence transformer that provides a high-dimensional latent space representation for each title \cite{reimers-2019-sentence-bert}. In Reddit, post titles often mirror the news headlines they reference, providing a contextual basis for estimating embeddings for the comments. We employ `all-distilroberta-v1'\footnote{The model is accessible from \url{https://huggingface.co/sentence-transformers/all-distilroberta-v1}} model for encoding post titles into a 768-dimensional array of real numbers. We chose this model for its high variance in cosine similarities across post titles in the dataset, a feature likely stemming from its extensive training on Reddit conversations \cite{henderson-2019-reddit-comments}.



\subsubsection{Stance Detection}
In the context of comment discussions, stance detection, or more specifically (dis)agreement detection, is the task of identifying the stance of a \textit{parent} text, usually the original post or a comment, to a \textit{child} text, which is a reply to the parent. we define the possible stances of one text to another using three discrete categories:
\begin{itemize}
    \item favor: The child text is supportive of the parent text.
    \item against: The child text opposes or otherwise criticizes the parent text.
    \item none: The child text is neutral towards the parent text.
\end{itemize}

We use the LLaMa-2-7b \cite{Touvron2023Llama2}, a large language text generation model with 7 billion parameters, to classify the stances of each comment towards its parent under the context of the post. We do this classification by running a text completion task on LLaMa-2-7b using the following prompt, which is inspired by \citet{anon2023stanceDetection}.

\begin{quote} 
\itshape
[INST]{\textless}SYS\textgreater You are a helpful, respectful, and honest assistant that detects the stance of a comment with respect to its parent. Stance detection is the process of determining whether the author of a comment is in support of or against a given parent. You are provided with:
post: the text you that is the root of discussion.
parent:  the text which the comment is a reply towards.
comment: text that you identify the stance from.

You will return the stance of the comment against the parent. Only return the stance against the parent and not the original post. Always answer from the possible options given below: 

favor: The comment has a positive or supportive attitude towards the post, either explicitly or implicitly.

against: The comment opposes or criticizes the post, either explicitly or implicitly. 

none: The comment is neutral or does not have a stance towards the post. 

unsure: It is not possible to make a decision based on the information at hand.{\textless}/SYS\textgreater

post: \{post\}

parent: \{parent\}

comment: \{comment\}

stance: [/INST]
\end{quote}

Here, the \textit{\{post\}}, \textit{\{comment\}}, and \textit{\{parent\}} represent the original post title, the comment to the post, and the parent to the comment, respectively. We limit the text completion to a maximum of $7$ new tokens, as this is the maximum amount of tokens required to return one of the three possible stance options. Then we compare the generated part of the text and match it to one of the three possible stances we have. 


Using the default model without any fine-tuning, the above approach performs very poorly at a classification accuracy approaching $0\%$. This is because there is no guarantee for the model even to generate text matching one of the options we require. 

To fine-tune the LLaMa-2-7b model, we use a Low-Rank Adaptation (LoRA) \cite{Hu2021Lora} \cite{anon2023stanceDetection}. LoRA introduces linear layers parallel to each layer in the LLaMa-2-7b model. These layers are matrices that linearly transform the input and add their output to the original layer outputs. Each weight layer of the LoRA is a matrix of rank $r$, which is a hyper-parameter we choose. This fine-tuning process effectively modifies the output of the text generation model while maintaining training efficiency. LoRA layers also have a second hyperparameter $\alpha$ which controls how much the output of the LoRA layers are scaled. Tuning this parameter can help increase the data efficiency of LoRA further. However, it is common practice to select $\alpha$ and $r$ equal to each other, and throughout the rest of this paper we always pick them the same.

In our experiments, we use a LoRA model of rank $r=16$ for fine-tuning the base LLaMa-2-7b model. We set the learning rate to $4\times10^{-4}$ and employ $10\%$ dropout \cite{Srivastava2014Dropout} during training. To train and validate the LoRA model we use the \dagg, which is an agreement disagreement dataset containing expert-labeled comment-reply pairs collected from Reddit. We provide more information on this dataset in the Data section of this paper. 

\subsubsection{Comment Embedding} 
The goal of embedding the comments is to identify the embedding of users in a similar SBERT latent space to the one we use in post-embedding. We achieve this by embedding the entire corpus of all comments of each user and then pooling them by averaging for each user to get embedding representations for the user. Intuitively, user embeddings capture the overall opinions, biases, and interests of users.

Unlike the original posts, comments and replies on social news websites rarely express complete statements or opinions on a specific topic by themselves. Thus, it is difficult to get latent embeddings of the opinion a comment expresses by directly using text embedding without using the original post as the context. As an illustration, consider the following example Reddit post-comment pairs.

\begin{quote}
\itshape
    \textbf{Post 1:} China will surpass US to be world's largest economy.
    
    \textbf{Comment 1:} valid point though.
    
    \textbf{Post 2:} Trump believed Comey intentionally misled the public to believe that he was under investigation 
    
    \textbf{Comment 2:} Completely valid point.
\end{quote}
Here, it is clear that comment 1 and comment 2 express vastly different opinions and biases despite nearly having the same textual content. Thus directly relying on comments to get user opinions does not work well without the original post providing context. We resolve this issue by using the original post embedding as a context and then assigning an embedding to each comment based on its stance towards the original post.



Suppose that we have a post $P$ with corresping embedding $h(P)$ obtained by encoding post title through SBERT. Now consider a comment $C$ to this post where the stance of the comment to post is denoted as $\sigma(C,P)$, where
\begin{equation}
    \sigma(C,P) = \begin{cases}
        1 &\quad \textrm{if } C \textrm{ favors } P, \\
        -1 &\quad \textrm{if } C \textrm{ is against } P, \\
        0 &\quad \textrm{if } C \textrm{ is neutral towards } P.
    \end{cases}
\end{equation}

Ideally, if the comment $C$ entails post $P$, their embeddings should be similar, and if the comment $C$ is against post $P$, its embedding should either be similar to the negation of $P$ if $P$ is a logical statement or contain a contradictory opinion to $P$. Thus letting $\neg P$ represent a negation to $P$ we define the embedding $h(C)$ of the comment $C$ as
\begin{equation}
    h(C) = \begin{cases}
        h(P) &\quad \textrm{if } \sigma(C,P) = 1, \\
        h(\neg P) &\quad \textrm{if } \sigma(C,P) = -1, \\
        \frac{h(P) + h(\neg P)}{2} &\quad \textrm{if } \sigma(C,P) = 0.
    \end{cases}
\end{equation}

The above assignment relies on knowing what the string $\neg P$ is, which is generally impossible. In fact, a rigorous definition for a negation string $\neg P$ might not exist. 
We overcome this issue by training a model to directly estimate $h(\neg P)$ from $h(P)$ on a dataset where negation strings are well-defined and known. Generalizing this model to arbitrary strings yields a model which transforms any given embedding into an embedding of a contrary opinion.


To prevent overfitting, we use a simple affine transformation as our negation model. That is, we find a matrix $\mathbf{A}$ and a bias vector $\mathbf{b}$ that transforms the embedding of a given string $S$ into the embedding of its negation $\neg S$ with minimal mean squared error loss $||\mathbf{A} h(S) + \mathbf{b} - h(\neg S)||_2$ over some well known negation dataset, and then \textit{stipulate} that for any post $P$ the embedding of its negation is $h(\neg P) = \mathbf{A} h(P) + \mathbf{b}$.

Note that in the ideal case where the input data consists only of strings that define a logical statements with a clearly defined negation, the affine transformation induced by $A$ and $b$ would be an \textit{affine involution}, that is, it would be its own inverse, as the negation of a negated statement must be the original statement. We do not enforce this constraint on our affine model as the sentence embeddings are not ideal, and unconstrained affine models can yield higher accuracy.

We use {\dneg}, which is a dataset containing sentence entailment and negation examples, to train the affine negation model parameters $\mathbf{A}$ and $\mathbf{b}$. We train the model both to transform the entailment examples to negation examples, and vice versa to ensure the model generalizes well to the negation of negative statements. Note that we use mean squared loss in fitting the affine model, instead of the cosine error, despite the latter being more common in language modeling tasks. Empirically we found the mean squared loss to provide better performance in later clustering steps as it limits the norm of the predicted $h(\neg P)$ embeddings to be small. After fitting the affine model the comments embeddings become
\begin{equation}
    h(C) = \begin{cases}
        h(P) &\quad \textrm{if } \sigma(C,P) = 1, \\
        \mathbf{A} h(P) + \mathbf{b} &\quad \textrm{if } \sigma(C,P) = -1, \\
        \frac{1}{2} ((\mathbf{A} + \mathbb{I}) h(P) + \mathbf{b}) &\quad \textrm{if } \sigma(C,P) = 0.
    \end{cases}
\end{equation}
where $\mathbb{I}$ denotes the identity matrix.
The \textit{user embeddings} follow directly from the comment embeddings by pooling all comments written by a user and averaging them over the time period of interest.

\subsection{Credibility and Political Bias Analysis}


We determine the credibility of each user by first assigning a credibility score, ranging from $0$ to $1$ to the original posts, then assigning scores to each comment based on the credibility of the parent post and the stance of the comment toward the post. Finally, we average the credibility of the comments of each user to get an average credibility score for each user. We use this score as a metric for how likely each user is to engage positively with low credibility content, with a lower score meaning higher susceptibility to misinformative sources. We follow the same steps to estimate the political bias of each user as well, with the only difference being that the bias scores range from $-1$ to $1$, denoting left-wing and right-wing political views respectively. \Cref{fig2}.b presents a visualization of the credibility and political bias assignment process.

\subsubsection{Post Credibility and Bias} We determine the credibility and biases of the posts using the credibility-bias rankings for news sites published by the data released by Ad Fontes Media Corporation. Approximately $29\%$ of the posts included in the four political subreddits in {\dred} include a reference to a verifiable news article. We then assign a credibility and bias rating for the post using the news article it references.

\subsubsection{Comment Credibility and Bias} We define a comment's credibility using the following equation
\begin{equation}
    \textrm{Cred}(C) = \sigma(C,P) \left(\textrm{Cred}(P) - \frac{1}{2}\right) + \frac{1}{2},
\end{equation}
where $\textrm{Cred}(C)$ and $\textrm{Cred}(P)$ denote the credibilities of the comment and its parent post, respectively. Notice that we rely on the user stances $\sigma(C,P)$ we derive from the fine-tuned LLM model described in the comment embedding section. That is to say, we assign the same credibility to the comment and post when the comment favors the post and assign credibility $1 -\textrm{Cred}(P)$ to the comment when it is against the post. We define the comment biases using a similar method, but as bias assignments are centered around $0$ we simply write
\begin{equation}
    \textrm{Bias}(C) = \sigma(C,P) \textrm{Bias}(P).
\end{equation}
Similarly to the case with comment embedding, we use pooling to derive user credibility/bias assignments from the comment averaging them for each user.

\section{Community Susceptibilities} 
After obtaining both user credibility and bias scores along with the user embeddings, we can analyze the credibility of user groups. While Reddit includes user-generated communities called subreddits, we avoid using these subreddits directly and instead follow a clustering-based approach to detect distinct interest groups. Subreddits have two main limitations that are problematic for determining groups of users with particular interests. Firstly, multiple user groups might exist in a single subreddit, creating separate competing cliques within each subreddit that do not agree with their interests. Perhaps the most stark example of this is the largest political discussion subreddit r/politics, where political views of all walks contribute and share news, and predictably discussions involving different political opinions are abundant. The second reason is that some users opinions might be tempted to follow and participate in multiple subreddits, for example, around half of all users in the subreddit r/Republicans also comment regularly in the subreddit r/Conservatives, yielding these subreddits ineffective in terms of partitioning users into distinct interest groups. 

Instead of using subreddits, we use \textit{spectral clustering} \cite{Yu2003Spectral, Damle2018Spectral} on user embeddings to detect interest groups. To better combat the noise in the user distributions, we adopt a local scaling method \cite{zelnik-manor2004stsclusterng}. In this method, we first compute the pairwise cosine distances 
\begin{equation}
    d(x,y) = \dfrac{\mathbf{x}^{\top}\mathbf{y}}{||\mathbf{x}||\cdot||\mathbf{y}||},
\end{equation}
of all user pairs $(x,y)$ with respective latent space embeddings $(\mathbf{x},\mathbf{y})$. We then define an affinity $\mathbf{W}$ using a Gaussian kernel as 
\begin{equation}
    \mathbf{W}_{x,y} = \exp \left( \frac{-d(x,y)^2}{\sigma_x\sigma_y} \right),
\end{equation}
where $\sigma_x, \sigma_y$ are the cosine distances to the $7$'th nearest neigbors of $x$ and $y$ respectively \cite{zelnik-manor2004stsclusterng}.

Spectral clustering of the latent space then simply becomes spectral clustering on the weighted graph with weighted adjacency $\mathbf{W}$ as described in \cite{Yu2003Spectral}. We determine the number of clusters to split the users into using the self-tuning spectral clustering approach described by \citeauthor{zelnik-manor2004stsclusterng} (\citeyear{zelnik-manor2004stsclusterng}), where the authors define an \textit{alignment score} to each number of clusters, and choose the number that yields minimal alignment cost.


Notice that these clusters are based solely on the user embedding, thus there is no explicit dependence between the cluster a user belongs to, and their credibility and bias score. We map these clusters onto the credibility bias space using the credibility and bias assignments of each user. We then analyze these distributions to estimate how positively each cluster reacts to high-bias or low-credibility news sources. 

\section{Data}
In this section we summarize the contents and the preparation of the datasets we use throughout the paper.
\subsection{\DRed}
{\renewcommand{\arraystretch}{1.15}
\begin{table}[t]
    \centering
    \begin{tabularx}{230pt}{c|Xccc}
        \hline
            \textbf{Year} & \textbf{Subreddit} & \textbf{\# Posts}& \textbf{\# Comments} \\ \hline 
            \multirow{4}{*}{2016} 
            &r/Conservative & 6242 & 23569 \\ 
            &r/Libertarian & 1792 & 5899 \\
            &r/Republican & 976 & 3680 \\
            &r/democrats & 2398 & 5553 \\ \hline
            \multirow{4}{*}{2017}
            &r/Conservative & 7358 & 28998 \\ 
            &r/Libertarian & 2169 & 7938 \\
            &r/Republican & 580 & 2240 \\
            &r/democrats & 830 & 2365 \\ \hline
            \multirow{4}{*}{2018}
            &r/Conservative & 11146 & 51646 \\ 
            &r/Libertarian & 3850 & 13739 \\
            &r/Republican & 400 & 1421 \\
            &r/democrats & 2047 & 5351 \\ \hline
    \end{tabularx}
    \caption{Post and comment numbers in the Reddit dataset used in the experiments.}
    \label{tab:redditData}
\end{table}}

This study uses real-world data collected from Reddit,\footnote{Accessible from \url{https://www.reddit.com}.
All data collected from \url{https://pushshift.io}.} which is a social news website based in the U.S. Reddit discussions are organized in broad, community-generated groups, called \textit{subreddits}, and consist of an original \textit{post}, followed by \textit{comments} and \textit{replies} to the comments. The data consists of the posts and the comments from three consecutive years, starting from January $2016$ and ending in December $2018$. We chose four subreddits to collect the data from: \textit{r/Conservative, r/Libertarian, r/Republican, r/democrats.} We chose these subreddits as they span a wide range of political biases. 

We prune the data by removing all deleted posts, and comments that contain less than three words. We also remove all users with less than $10$ comments in any given year, as these users have too few comments to reliably assign them a latent-space representation. We also remove all of the posts that do not contain any comments after pruning as they do not contain any information on users. \Cref{tab:redditData} shows the distribution of posts and comments after pruning. The entire corpus of the comments in this pruned data are authored by $3,155$ users, providing an ample source of comments per user to achieve accurate clustering.

\subsection{\DAgg}
To train the LoRA model, we use an agreement/disagreement dataset \cite{pougue-biyong2021debagreement}, which contains $42,894$ comment-reply pairs with expert annotated stances between each pair. We generate training and validation sets by first removing all comment-reply pairs that are unlabeled or have conflicting labels between multiple experts. This pruning effectively filters out most of the outliers in the data, which is required as LoRA models are often susceptible to outlier-caused performance losses, a known side effect of their high data efficiency. Next, we sample $10,000$ of these comments-reply pairs and split it into two partitions of $9,000$ and $1,000$ corresponding to training and validation sets respectively.
\subsection{\DNeg}
The negation dataset \cite{günther2023jina} consists of $10,000$ sentence triplets. Each triplet consists of the following strings:
\begin{itemize}
    \item Anchor: A base string,
    \item Entailment: A string that follows logically from the anchor,
    \item Negation: A string that contradicts the anchor.
\end{itemize}

The dataset is based on the SNLI dataset \cite{snli:emnlp2015}, which is composed of sentence pairs, consisting of an anchor and a hypothesis, with human generated labels describing whether the hypothesis logically follows the anchor, or contradicts the anchor. The entailment and negation strings come directly from the positively and negatively labeled hypotheses that share a common anchor. We embed all of these sentences using the `all-distilroberta-v1' model. We then split the available sentences to a $9,000$ training samples and $1000$ validation samples.

Naturally, the entailment and negation strings are negations of each other. When training the comment embedding layer, we thus inflate this dataset by constructing all possible tuples in the form of [entailment, negation], or [negation, entailment], yielding $20,000$ tuples of sentences that contradict each other, yielding a total of $18,000$ training samples and $2,000$ validation samples

\section{Results}

We first present the training and validation results of the LoRA fine-tuning model for stance detection, and the affine transformation model for negation embeddings. We then present our main results.

\subsection{LoRA Fine-Tuned Stance Detection}
LoRA fine-tuning of the base LLaMa-2-7b model contributes to a dramatic increase in the accuracy of the stance detection. The majority of this performance increase is due to the fine-tuned model being much less averse to returning text completions that are outside the three allowed categories, with a slight and gradual increase in performance in later epochs due to the fine-tuned model becoming more capable at identifying language nuances in Reddit comment discussions. \Cref{Table:lora results} summarizes the classification accuracy of the fine-tuned and the base model, along with some other LoRA training parameters we have tested. These results show performances that are comparable with the reported state-of-the-art results for stance detection tasks using {\dagg} \cite{luo2023daggresult} which reports a mean F1 score of $66.91\%$.

{\renewcommand{\arraystretch}{1.15}
\begin{table}[t]
    \centering
    \begin{tabularx}{220pt}{Xccc}
        \hline
        \textbf{Model} \hspace{10mm}& \textbf{Train Acc.} & \textbf{Val. Acc.} & \textbf{F1} \\
        
        \hline
        Base Model & - &  $0\%$& $0$  \\
        Rank 4 LoRA & $65\%$ & $61\%$ &  ${58\%}$\\
        Rank 8 LoRA & $68\%$ &  $65\%$ & ${63\%}$\\
        \textbf{Rank 16 LoRA} & $\mathbf{73\%}$ & $\mathbf{68\%}$ &  $\mathbf{66\%}$\\
        Rank 32 LoRA & $74\%$ & $67\%$ & ${65\%}$\\
        \hline
    \end{tabularx}
    \caption{Comparison of model accuracies of the base LLaMa-2-7b model and four fine-tuned versions. We use the model indicated in bold for the rest of the experiments. }\label{Table:lora results}
\end{table}}

The entire LoRA fine-tuning took $2$ hours and $45$ seconds running on a single NVIDIA RTX A5000 GPU.

\begin{figure}[t]
\centering
\includegraphics[width=\linewidth]{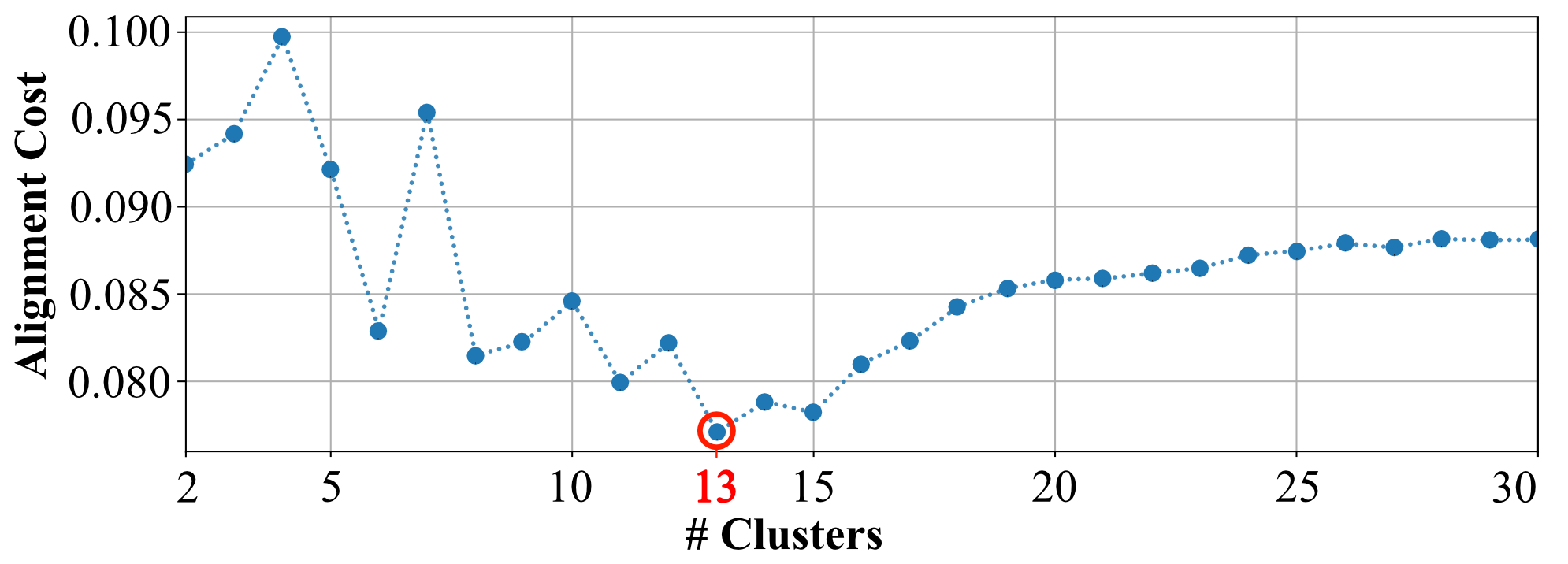} 
\caption{Alignment cost across different numbers of clusters for users' latent space embeddings. }
\label{fig:alignment}
\end{figure}

\subsection{Affine Negation Model}
We train the affine model for sentence embedding negation using {\dneg} as explained in the methods section. It took $31$ epochs for the affine model to reach minimal mean squared loss across the validation set. The training results yielded a mean squared loss of $4.11\times 10^{-4}$ and the validation results yielded a loss of $4.88\times 10^{-4}$. The training took approximately $5$ minutes and $30$ seconds running on a single NVIDIA RTX A5000 GPU.

To verify the effectiveness of the affine negation model, we also compare the cosine similarity between the ground truth negation embeddings in the validation set and the predicted negation embeddings returned by the negation model. We discover that there is an average cosine similarity of $0.79$ between the predicted embedding and the ground truth. We contrast this error with the naive approach of inverting the sign of the embedding for negation, which yields a cosine similarity of $-0.12$, a significant decrease in performance compared to using the affine negation model considering that a cosine similarity of $0$ indicates that the ground truth embeddings and predicted embeddings are orthogonal.

\subsection{Main Results}


We investigate the variation in the susceptibility to news sources of differing credibility among Reddit user communities. Results reveal marked differences in how these clusters respond to biases and credibility in news sources (\textbf{RQ1}). Specifically, in some clusters, susceptibility to low-credibility news is as much as three times higher than in others. This finding indicates that cluster association is indicative of users' susceptibility to low-credibility news (\textbf{RQ2}). The susceptibility of users against biased media shows similar, trends with some communities having significantly more highly biased users than others. These insights underscore the importance of cluster-specific strategies in combating low-credibility and highly biased news propagation. 

\subsubsection{User Clustering}

We cluster users based on their latent space embeddings using self-tuning spectral clustering, a technique that captures the underlying patterns in user interactions with posts \cite{zelnik-manor2004stsclusterng}. To identify the optimal number of clusters, we calculate the alignment score for each potential cluster number, ranging from $2$ to $30$. \Cref{fig:alignment} shows alignment scores as a function of the number of clusters, showing that $13$ clusters attain the minimal alignment cost, indicating the optimal number of clusters. We find that using $4$ clusters, which corresponds to the number of subreddits in our study, attains the highest alignment cost. This disparity suggests that subreddit categories alone do not provide a meaningful way to cluster users based on their interaction patterns with posts. Thus, our analysis justifies the decision to not rely on subreddit categorization for defining user communities


\begin{figure*}[t]
\centering
\includegraphics[width=\linewidth]{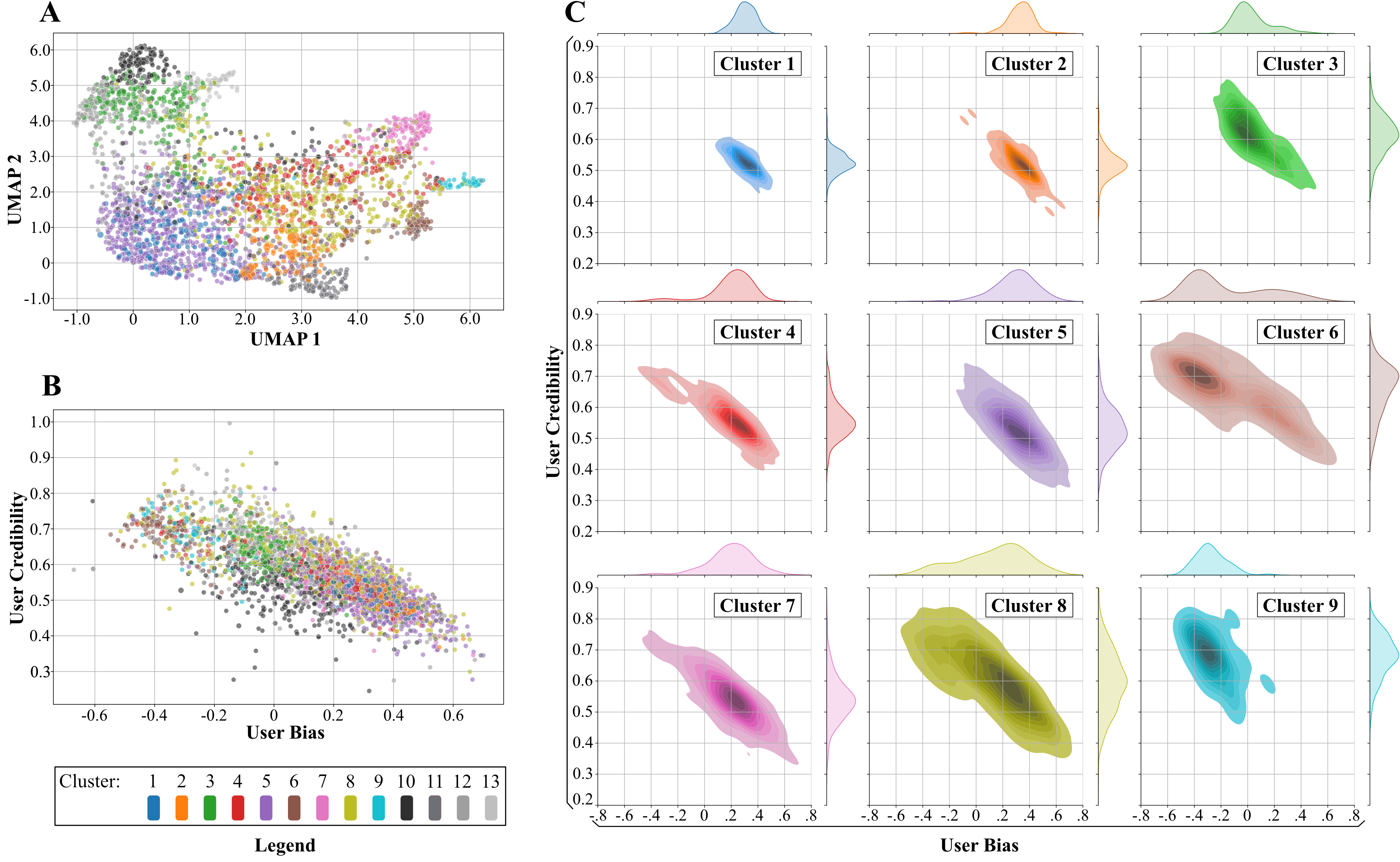} 
\caption{User distributions of all $13$ clusters across years 2016-2018. \textbf{A:} Latent space embedding visualization of users using UMAP reduction. \textbf{B:} Credibility-bias mappings of all users. Larger numbers denote higher credibility and right-wing political bias in their respective axes. \textbf{C:} Credibility-bias distributions of $9$ clusters with least maximal covariance eigenvalue, together with marginal distributions.}
\label{fig3}
\end{figure*}

{\renewcommand{\arraystretch}{1.20}
\begin{table*}[t]
    \centering
    \begin{tabularx}{500pt}{ccX|ssss|cc}
    \hline
        
    \textbf{Cluster}& \textbf{\# Users}& \textbf{Dom. Subreddit} & \textbf{Mean Bias} & \textbf{Std. Bias} & \textbf{Mean Cred.}& \textbf{Std. Cred.} & $\mathbf{|\textrm{Bias}| > 0.5}$ & $\mathbf{\textrm{Cred.} < 0.5}$ \\

    \hline 
    1 &164 & r/Conservative& 0.305 & 0.076 & 0.526 & 0.033 & 00.0\% & 18.3\% \\
    2 & 112& r/Conservative& 0.331 & 0.099 & 0.519 & 0.045& 02.7\% & 29.5\% \\
    3 &232&  r/Libertarian& 0.032 & 0.138 & 0.614 & 0.063 & 00.4\% & 04.3\% \\
    4 &194& r/Conservative& 0.197 & 0.173 & 0.558 & 0.057 & 00.5\% & 12.9\% \\
    5 &674& r/Conservative& 0.292 & 0.164 & 0.528 & 0.071 & 08.0\% & 34.7\% \\
    6 &168& r/democrats& -0.152 & 0.297 & 0.655 & 0.078 & 03.6\% & 04.8\% \\
    7 &155& r/Conservative& 0.198 & 0.176 & 0.542 & 0.071 & 02.6\% & 25.8\% \\
    8 &520& r/Conservative& 0.127 & 0.257 & 0.601 & 0.098  & 05.6\% & 16.0\% \\
    9 &61& r/democrats& -0.271 & 0.116 & 0.680 & 0.065 & 00.0\% & 01.6\% \\
    10 &291& r/Libertarian& 0.021 & 0.155 & 0.531 & 0.079 & 00.3\% & 32.6\%\\
    11 &169& r/Conservative& 0.286 & 0.220 & 0.532 & 0.078 & 08.3\% & 34.3\% \\
    12 &308& r/Libertarian& 0.055 & 0.205 & 0.622 & 0.100 & 03.2\% & 10.7\% \\
    13 &107& r/Libertarian& -0.001 & 0.142 & 0.641 & 0.077 & 00.0\% & 03.7\% \\
    \hline
    \end{tabularx}
    \caption{Comparison of credibility and bias distributions of clusters.}\label{Table:clusters_stats}
\end{table*}}

We employ \textit{uniform manifold approximation and projection} (UMAP) \cite{mcinnes2018umap-software} to visualize the distribution of the $13$ user clusters in a reduced two-dimensional space. UMAP helps in simplifying complex, high-dimensional user data into a format that's easier to visualize. \Cref{fig3}A shows the distribution of the users in a two-dimensional UMAP representation. This color-coded UMAP representation demonstrates distinct user communities, visibly separated in the two-dimensional space. This separation serves as visual evidence of the spectral clustering method's effectiveness in categorizing users into discrete and meaningful groups. The UMAP plot also reveals patterns in user behavior, such as the concentration of certain clusters, which warrants further investigation.


The descriptive analysis of clusters reveals significant variations in size and spread. Of the total population of $3,155$ users, $61$ are in the smallest cluster and $674$ belong to the largest cluster. The populations of other clusters roughly distributed according to a power law. We measure the spread each cluster occupies in the embedding space by calculating the principal component standard deviation (PC-std), the greatest standard deviation of a cluster across all possible directions. Calculating PC-std equates to finding the square root of the largest eigenvalue of the covariance matrix of each cluster. The PC-std values ranged from $0.036$ in the most compact cluster to $0.100$ in the most dispersed one, indicating a nearly threefold variation in cluster spans. We sort the clusters based on increasing order of PC-std, meaning cluster $1$ is the cluster with the tightest distribution and cluster $13$ has the widest distribution. Note that \Cref{fig3}A does not represent the spread of these clusters accurately due to the non-linear projection of UMAP, for example, cluster $9$, which is the fourth most spread out cluster in the embedding space, appears tightly distributed in \Cref{fig3}A.


\subsubsection{Correlation to Credibility and Bias}
\Cref{fig3}B shows the projection of the user embedding onto the source credibility and political bias space where each color represents a different user cluster. Notice that despite there being no explicit dependence between cluster assignments and user credibility and bias scores, there is a visible separation between the distributions of each cluster. We also visualize the individual cluster distributions in \Cref{fig3}C and summarize their distributional characteristics in \Cref{Table:clusters_stats}. 

The results indicate a notable correlation between political bias and credibility with a Pearson's correlation constant of $-0.76$ among users on social news websites. Specifically, users with left-leaning tendencies tend to have higher credibility scores on average. This correlation contradicts the news source distributions in the Ad Fontes Media dataset. We discuss the implications of this representation further in the discussion section.

A significant observation is that overall, the clusters that have tighter distributions in the latent space also have a tight distribution in the credibility bias space. Indeed, we find a Pearson's correlation of $0.41$ between the PC-stds of the latent space embeddings and the credibility bias embedding, meaning that users that have similar latent space embeddings also tend to have similar credibility and bias scores. 
The significance of this finding lies in the independent nature of the credibility-bias assignment from the latent space embeddings. 
This correlation shows that some latent space features are indicative of susceptibility to highly biased or uncredible content.

To understand how these clusters align with Reddit subreddits, we examine their \textit{dominant subreddit}---the subreddit where users in a cluster most frequently post comments. We find that none of the $13$ clusters have r/Republican as their dominant subreddit, mainly due to the total number of comments from r/Republicans being small compared to other subreddits. There is a major variability between clusters in the proportion of comments included in the dominant subreddit. $99.2\%$ of comments made by users from cluster $9$ is in r/democrats, the dominant subreddit of cluster $9$. In contrast, this ratio is only $51.0\%$ for cluster 10. The ratio is between these two extremes for other clusters. \Cref{Table:clusters_stats} includes the dominant subreddits for all clusters.

We determine the threshold for low credibility users as those with a credibility score less than $0.5$; likewise, we determine the highly biased users as those with bias greater than $0.5$ or less than $-0.5$. These thresholds correspond to low credibility and hyper-partisanship thresholds in the Ad Fontes Media Dataset. Overall, the results show that cluster associations strongly impact users' susceptibility to uncredible and highly biased news. Cluster $5$ and cluster $9$ achieve the highest and the lowest proportion of low credibility users at $34.7\%$ and $1.6\%$ respectively. This difference is significant as it means a member of cluster $5$ is over $20$ times more likely to be a low credibility user than a user in cluster $9$. This ratio comparison is not possible for comparing the proportion of highly biased users as three of the $13$ clusters have $0$ that are highly biased. \Cref{Table:clusters_stats} presents the proportion of highly biased and low credibility users across all clusters.

Our work introduces a credibility/bias score that proves instrumental in identifying echo chambers. We can vividly visualize echo chambers within user clusters by measuring users' tendencies toward biased sources. For instance, clusters $1$, $2$, and $9$ exhibit features characteristic of echo chambers, such as a high degree of tight spread in bias and predominant subreddits. Conversely, clusters like $6$ and $8$ demonstrate a broader range of spread in terms of bias, indicating a more diverse mix of opinions.

Our findings align with those of Morini et al. \citeyearpar{morini2021toward}, who explored echo chambers' existence and temporal dynamics on specific topics. This is similar to our observations in clusters $1$, $2$, $5$, and $9$. Meanwhile, De Francisci Morales et al. \citeyearpar{de2021no} challenge the ubiquity of echo chambers in certain political discussions, a notion supported by our analysis of clusters $3$, $6$, and $8$. The strength of our approach lies in its foundation on embeddings derived from the posts. This enables our model to distinguish not only users' biases and reactions but also their responses to different topics. This provides a significant leap towards automating such analyses, allowing for a more nuanced and detailed study of user behavior in online communities by not relying on users' comments but rather on their reactions to topics.

\section{Discussion}
In this section we interpret and discuss the results of our analysis. We provide the major limitations and discuss avenues for future work. 

\subsection{Implications of Dataset Bias} 
The {\dred} features a significant portion of posts from r/Conservatives. This subreddit predominantly features right-leaning content, with over $0.91\%$ of its shared news originating from right-leaning sources. We explore the implications of this bias in the following discussion.

A key observation we made in the results section is that there is a correlation between political bias and credibility, where right-leaning users also scored lower in their credibility assignment. This correlation is not necessarily indicative of a correlation in the news sources. Contrarily, the credibility and bias distribution of the news sources in the Ad Fontes Media dataset show that both the extreme right and extreme left news sources are associated with low credibility in a similar fashion. We hypothesize that this discrepancy between user credibility and source credibility may stem from the predominant sharing of extreme right sources over extreme left ones in the subreddits under our study. This imbalance causes right-leaning posts to have a lower credibility than left-leaning posts on average in our dataset, which likewise affects the user credibility.

Our analysis of the distribution of clusters also show that most of the clusters have a mean bias score leaning towards right-wing politics. In addition, observing \Cref{fig3}C, there are some user clusters, such as cluster 6, that have a distribution that contains both highly left-leaning and highly right-leaning users. This is likely due to incorrectly clustered right-leaning users presenting as noise in cluster 6 which otherwise mainly contains left-leaning users.

Despite these issues, the user clustering in the latent space admits meaningful separation of right-leaning and left-leaning users. This is mainly thanks to the local scaling step we use in the spectral clustering method, which separates tightly packed clusters from the more spread-out background noise caused by incorrectly embedded users.

\subsection{Limitations}
Our study faces two major limitations concerning the validity of our results. First, the task of determining the credibility of news sources is inherently complex and somewhat subjective. Relying on a single source for credibility and bias analysis constrains the validity of our conclusions. Second, our method for defining user credibility and biases involves an implicit assumption: users who regularly react negatively to high-credibility content, as opposed to low-credibility content, are considered to have lower credibility. This assumption is not completely unjustified, as users who consistently respond negatively to high-credibility content, in contrast to low-credibility content, demonstrate a discernible preference for the latter. However, it overlooks th additional reasons for a highly credible user to reach negatively to a highly credible news source such as conflicting political views. These considerations motivates future work aimed to detect user credibilities more accurately.

\subsection{Future Work}

Future work can expand our analysis in two key directions. Firstly, to improve the validity of our results, we suggest expanding the dataset and refining the methods for assessing user credibility and political biases. Advances in stance detection and sentence embedding methods can lead to more accurate user embeddings. These advancements could produce clusters with tighter distributions, allowing more granular analysis of their credibility and bias distributions. The second major direction for future work is to evolve our proposed user embedding pipeline to include more nuanced effects such as integrating the content of user comments alongside their stances towards posts. Additionally, employing graph-based methods to capture the interactions among comments could further refine the embeddings and consequently reveal richer conclusions on the credibility and bias susceptibilities of user clusters.

\section{Conclusion}
This paper introduces a novel pipeline to derive latent space embeddings of users from the sentence analysis of posts and comments in Reddit. We show that this embedding pipeline induces a clustering of users into communities that have distinctive susceptibilities to incredible and highly biased news sources. Our experiments demonstrate that clusters that are tightly distributed in the embedding space tend to have a tight distribution in the credibility bias space, indicating that the user embeddings we derive are indicative of the credibility and bias scores of the users. Additionally, our research demonstrates that these user-generated communities do not inherently produce natural clusters in the latent space embeddings. This observation suggests that participation in subreddits does not necessarily mirror users' opinions or their responses to specific subjects.

\subsection{Privacy Considerations}
All of the experiments and methods in this paper use publicly available data. We do not disclose any personal information in a manner that jeopardizes anonymity. We did not collect personal information on users other than the information available publicly from preexisting datasets, and we anonymized all sample messages and discourse that we show explicitly in this paper. This paper was not subject to the academic IRB process.

\bibliography{references.bib}
\end{document}